\pdfoutput=1
\documentclass[aip, jcp, 
amsmath, amssymb,
reprint
]{revtex4-1}
\usepackage{graphicx}
\usepackage{xspace}

\usepackage{lmodern}

\usepackage[T1]{fontenc} 
\usepackage[utf8]{inputenc}
\usepackage{esint}
\usepackage{bm}

\newcommand{\Avg}[1]{\langle #1\rangle}
\newcommand{\oX}{{\bar{X}}}
\renewcommand{\d}{\ensuremath{\mathrm d}}
\newcommand{\ts}[1]{\ensuremath{_{\mathrm{#1}}}}
\renewcommand{\vec}[1]{\ensuremath{\mathbf{#1}}}
\newcommand{\mat}[1]{\ensuremath{\bm{\mathsf{#1}}}}

\usepackage[american]{babel}
\begin{document}

\date{\today}

\title{Rare switching events in non-stationary systems}
\author{Nils B. Becker}
\affiliation{FOM Institute for Atomic and Molecular Physics (AMOLF),
 Science Park 104, 1098 XG Amsterdam, The Netherlands}
\author{Pieter Rein ten Wolde}
\affiliation{FOM Institute for Atomic and Molecular Physics (AMOLF),
 Science Park 104, 1098 XG Amsterdam, The Netherlands}

\begin{abstract}
Physical systems with many degrees of freedom can often be understood in
terms of transitions between a small number of metastable states.
For time-homogeneous systems with short-term memory these transitions are fully
characterized by a set of rate constants. We consider
the question how to extend such a coarse-grained description to
non-stationary systems and to systems with finite memory. We identify the
physical regimes in which time-dependent rates are meaningful, and state
microscopic expressions that can be used to measure both externally
time-dependent and history-dependent rates in microscopic simulations.
\end{abstract}

\maketitle

\section{Introduction}\label{sec:Intro}

Physical systems with a large number of degrees of freedom and complicated
dynamics can be understood in much simpler terms, if they exhibit a small
number of metastable states. A natural coarse-grained description is then
found in terms of residence times in each of the states, joined by
nearly instantaneous transitions between them.

This idea has been widely used in the context of thermal equilibrium systems
that are characterized by a clear separation of time scales between fast
intra-state dynamics and slow global relaxation (e.g.~\cite{haenggi90}), giving
rise to Markov State Models, see e.g.~\cite{prinz11}. Here
transitions are Poissonian, i.e.~they happen with uniform propensity
(probability per unit time); equivalently, the waiting time intervals are
uncorrelated and exponentially distributed. Each transition is described by a
single number, the rate constant.

In this article we are interested in systems that switch between two metastable
states rarely and rapidly, but with a time-dependent propensity.
Such time-dependence may be caused by an external force driving the system.
Alternatively it may also arise due to the presence of internal degrees of
freedom that do not fully equilibrate on the macroscopic time scale of
interest, giving rise to non-Markovian macroscopic switching dynamics between
the metastable states. We consider both equilibrium and non-equilibrium
systems, in the sense that their unperturbed microscopic dynamics may or may
not be time-reversible.

Clearly, in such time-inhomogeneous or non-Markovian systems the
concept of a \emph{rate constant} is inadequate. In this 
article we investigate in what physical regimes a
time-dependent \emph{rate function} or a history-dependent \emph{rate kernel}
may instead be meaningfully defined, and state the corresponding
phenomenological rate equations. 
We then address the question how macroscopic rate
functions or rate kernels can be defined in terms of microscopic
correlation functions; we give microscopic expressions and show how they can be
used to measure these generalized rates in computer simulations.

\section{Overview}\label{sec:Metastable}

We consider systems whose coarse-grained, macroscopic dynamics can be described
as switching between two macroscopic states $A$ and $B$.   The macroscopic
states are assumed to be metastable in the sense that the mean waiting time
between the switching events is much longer than the duration of the switching
event itself. This makes it possible to partition the phase space into the
metastable regions $A$ and $B$, and a transition region $C$ separating them,
such that
\begin{equation} 
\tau_C \ll \tau_{AB},\tau_{BA},
\label{eq:tau_C}
\end{equation}
where $\tau_C$ is a typical duration for a traversal of $C$ (a switching
event), and $\tau_{AB}, \tau_{BA}$ are the mean waiting times for switching in
the forward and backward direction, respectively.  This separation of time
scales provides the justification for coarse graining the system as a two-state
system, switching rarely but rapidly between the states $A$ and $B$.

We are interested both in equilibrium systems that are microscopically
reversible and in non-equilibrium systems with dissipation of energy.
If the system is in stationary state, the stability of the states $A$ and $B$
and the forward and backward probability fluxes, $q_{AB}$ and $q_{BA}$, are
constant in time. Nonetheless, if the system has memory, the switching
propensities will be history-dependent. Furthermore, out of stationary state,
the stability of the states and the dynamics of switching between them will
change in time. In both of these situations, rate constants $k_{AB}$ and
$k_{BA}$ do not adequately describe the system, and a time-dependent
generalization of the rate constant concept is called for.

Following Chandler in his derivation of rate constants in equilibrium systems
\cite{chandler78}, we imagine that from experimental observations we know a
certain phenomenological expression for the macroscopic switching dynamics to
be valid, on times much larger than a certain macroscopic time resolution
$\Delta t$.
This macroscopic expression is the rate equation. The goal is then to derive a
microscopic expression for the dynamics of the system that is consistent with
the rate equation, and allows us to measure the macroscopic rates in a
microscopic simulation. We will thus not derive a macroscopic expression from
microscopic principles; rather, we will \emph{assume} that the system obeys a
given rate equation, and use this description as a starting point for the
derivation of microscopic expressions for time-dependent rate functions
\cite{chandler78}.

The propensity of the system to switch from one macroscopic state to another at
time $t$ may depend on the current macroscopic state of the system only, in
which case the system is Markovian. More generally, it may depend also upon the
macroscopic history, i.e.~the sequence of states visited in the past. For such
a non-Markovian system, the switching propensity could depend on the time that
has passed since the last switching event, but it is also conceivable that it
depends on the system's dynamics prior to the last switching event.
Indeed, all information about the history of the macroscopic switching dynamics
is contained in the sequence of switching times. It is thus natural to capture
the history dependence of the switching dynamics by writing the time evolution
of the system in terms of the times of all previous switching events:
\begin{eqnarray}
\frac{\partial}{\partial t} P_B(t;t',t'',\dots)
&=&k_{AB}(t|t',t'',\dots)
  P_A(t;t',t'',\dots) \nonumber \\
&& -
  k_{BA}(t|t',t'',\dots)
  P_B(t;t',t'',\dots), \nonumber\\
&&\label{eq:ME}
\end{eqnarray}
In this Master equation, $P_A(t;t',t'',\dots)$ denotes the \emph{joint}
probability to be in $A$ at time $t$ and to have switched for the last time
within $(t',t'+\d t')$ (i.e. from $B$ to $A$), for the second-to-last time in
$(t'', t''+\d t'')$ (from $A$ to $B$), and so on. Moreover,
$k_{AB}(t|t',t'',\dots)\d t$ is the conditional probability to leave $A$ in the
time interval ($t,t+\d t$) given that the sequence of switching times was
$t',t'',\dots$. An analogous equation holds with $A$ and $B$ interchanged.
Importantly, this macroscopic phenomenological rate equation \emph{defines} the
general rate  kernels $k_{AB}(t|t',t'',\dots)$ and $k_{BA}(t|t',t'',\dots)$.
An equation for the switching dynamics at time $t$ can
be obtained from Eq.~\ref{eq:ME} by integrating over the switching times prior
to $t$, \begin{equation}\label{eq:MEl}
\frac {\d}{\d t}P_B(t)= \int_{t>t'>t''>\dots}\frac{\partial}{\partial t}
P_B(t;t',t'',\dots)\d t' \d t''\dots.
\end{equation}
Since the last, second-to-last, $\dots$ switches are each unique
events, no over-counting occurs here. No approximation has
been made up to this point; note however that \ref{eq:MEl} is not in 
general a closed equation for $P_{A,B}(t)$. The lower integration limit
depends upon the experimental setup and will be discussed in more detail below.

While Eqs.~\ref{eq:ME} and \ref{eq:MEl} describe the switching dynamics of an
arbitrary two-state system, they cannot be solved in general. They are also too
detailed, requiring full information about the macroscopic history. At this
point the experiment has to inform us about the most useful and meaningful
phenomenological model, that is, the minimal model that captures the
macroscopic switching dynamics of the system. More specifically, the experiment
has to reveal whether the switching propensities do indeed depend on the
current time $t$ and the previous switching times $t', t'', \dots$.
The decision about the type of macroscopic expression that
describes the experiment best, will then depend both on the physical properties
of the system under consideration, and on the time resolution and the
measurement uncertainty of the experiment. Below, we will discuss a number of
stationary and non-stationary situations for which Eqs.~\ref{eq:ME} and
\ref{eq:MEl} can be simplified, and for which microscopic expressions for the
macroscopic rate constants can be found.

We now turn to the microscopic characterization of the system. We consider an
ensemble of trajectories that start at $t=0$ at state space points $x_0$,
following a specified initial phase-space distribution function $\rho(x_0)$
\footnote{If the microscopic dynamics is not Markovian, the initial condition
will have to be supplemented with a prescription of how the system was prepared
at $t<0$.}.
If the dynamics obey detailed balance and microscopic reversibility, and if
$\rho(x_0)$ is the canonical distribution, then this is an equilibrium
ensemble. We may instead also consider the relaxation towards thermodynamic
equilibrium, starting from a non-equilibrium $\rho(x_0)$. Alternatively, the
ensemble could be that of a non-equilibrium system that does not obey detailed
balance and microscopic reversibility. Again, the initial condition could
either be in stationary state, or out of stationary state, in which case the
non-equilibrium system may relax back to a stationary state. Finally,  the
system (time-reversible or not) may never reach a stationary state over the
time course $[0,T]$ of the experiment. This can happen if relaxation is slower
than $T$, or if the system is driven externally via some protocol $\phi(t)$,
for instance, proteins that are unfolded under the influence of an external
force.

To derive microscopic expressions that are consistent with the macroscopic rate
equations, Eqs.~\ref{eq:ME} and \ref{eq:MEl}, we first need to define functions
$h_A(x_t)$ and $h_B(x_t)$ that indicate whether the system with configuration
$x_t$ at time $t$ is in state $A$ or $B$. It is useful to define these
characteristic functions in terms of an order parameter $q(x_t)$ that serves to
measure the progress of the transition between the states $A$ and $B$:
\begin{eqnarray}
h_A(x_t) &=& \theta[q_A - q(x_t)],\\
h_B(x_t) &=& \theta[q(x_t) - q_B].
\end{eqnarray}
Here, $\theta$ is the Heaviside step function. With these definitions, the
system is considered to be in $A$ when $q(x_t) < q_A$, in $B$ when
$q(x_t)>q_B$, and in the transition region $C$ otherwise. Moreover,
we have the relation
\begin{equation}
h_A + h_B\equiv 1 \text{, if }q_A=q_B=q^*.\label{eq:unity} 
\end{equation}
The same relation holds effectively if $q_A<q_B$ and the occupancy of 
the transition region $C$ is low for all times.

The principal idea is now that the macroscopic rates $k_{AB,BA}(t)$ can be
derived from the behavior of the microscopic correlation function
\begin{eqnarray}\label{eq:C_t}
C(t)&=&\int \d x_0 \d x_t\rho(x_0) p(x_t|x_0)h_B(x_t)\nonumber\\
&=&\Avg{h_B(x_t)}
\end{eqnarray}
Here, $p(x_t|x_0)$ is the probability that the system is in state
$x_t$ at time $t$ given that it started in state $x_0$ at time zero;
$\Avg{\dots}$ denotes an average over the ensemble of trajectories
that start in $x_0$, following the phase-space distribution
$\rho(x_0)$. The correlation function $C(t)$ gives the probability
that the system is in state $B$ at time $t$ given that it
started from the initial distribution $\rho(x_0)$, which may or may not be
chosen to fully reside in $A$.  The transition rates are
derived from the flux into $B$ and out of $B$, and are thus related to
the time derivative of $C(t)$:
\begin{equation}
\dot{C}(t)=\Avg{\dot{h}_B(t)},\label{eq:Cdot_0}
\end{equation}
where we use the shorthand notation $h_B(t)=h_B(x_t)$.
The task at hand is to rewrite this microscopic expression such that it can
be identified with the macroscopic expressions of Eqs.~\ref{eq:ME} and
\ref{eq:MEl}. This is carried out below for a number of different classes
of systems. In each case, we start with the macroscopic rate equation,
and then derive a microscopic correlation function from which the macroscopic
rates can be obtained.

\section{Markov systems}\label{sec:StatSys}

In this section we consider systems that exhibit Markovian macroscopic
switching dynamics. By this we mean that it is known from experiment
that there exists a macroscopic time resolution $\Delta t$, on which the
propensity to switch between the two macroscopic states $A$ and $B$ at time $t$
is independent of the history of the macroscopic switching dynamics for earlier
times $t'<t$. In this case the rate kernels are independent of the previous
switching times, yet may depend on the current time if the system is not
time-homogeneous:
$k_{X\oX}(t|t',t'',\dots)=k_{X\oX}(t)$, where $X\oX=AB$ or $BA$. We immediately
conclude that the macroscopic rate equations Eqs.~\ref{eq:ME}, \ref{eq:MEl}
reduce to
\begin{equation}\label{eq:dPBmarkov}
\frac{\d} {\d t}P_B(t)= k_{AB}(t)P_A(t) -
k_{BA}(t)P_B(t)
\end{equation}
with rate functions $k_{AB}(t), k_{BA}(t)$. This equation together with
its counterpart for $P_A$ forms a closed set of equations for $P_{A,B}$. 
We remark that the form of Eq.~\ref{eq:dPBmarkov} can always be
obtained trivially from Eqs.~\ref{eq:ME}, \ref{eq:MEl} by
marginalizing. However, the resulting expression will then in general
only be valid for a particular experiment. A Markov system is
characterized by the property that Eq.~\ref{eq:dPBmarkov} with a
\emph{fixed} set of rate functions $k_{X\oX}(t)$ correctly describes
multiple experiments, differing in their initial conditions and in their
history before $t=0$.

The fact that memory is lost on the macroscopic time scale $\Delta t$
implies that the system locally equilibrates in the macroscopic states $A$ and
$B$ on time scales $\tau_A$ and $\tau_B$, respectively, that are shorter than
$\Delta t$. We can then identify a transient time scale 
\begin{equation}
\tau\ts{trans}=\max\{\tau_A,\tau_B,\tau_C\}<\Delta t \ll
\min\{\tau_{AB},\tau_{BA}\}
\label{eq:tau_A_B}
\end{equation}
which is the microscopic memory time of the system.

It is instructive to restrict the attention to trajectories that happen to
start in $A$ at $t=0$. To derive microscopic expressions for the macroscopic
rate constants, we then consider the correlation function
\begin{equation}
\dot{C}(t)=\Avg{\dot{h}_B(t)}_{A_0},\label{eq:Cdot_0_0}
\end{equation}
where the subscript $A_0$ indicates restriction of $\rho(x_0)$ to this
subensemble.
We now rewrite this expression  such that its terms can be
identified with the macroscopic quantities appearing in
Eq.~\ref{eq:dPBmarkov}. Taking $q_A=q_B=q^*$ and letting $h_X(t) = h_X(x_t)$,
we can insert $h_A(t-\Delta t) + h_B(t-\Delta t)=1$ and 
$h_A(t+\Delta t) + h_B(t+\Delta t)=1$ into Eq.~\ref{eq:Cdot_0_0}; the
importance of $\Delta t$ will become clear shortly. This yields
\begin{eqnarray}
\dot{C}(t)&=&\Avg{h_A(t-\Delta t)\dot{h}_B(t)h_A(t+\Delta t)}_{A_0}\nonumber\\
&+& \Avg{h_A(t-\Delta t)\dot{h}_B(t)h_B(t+\Delta t)}_{A_0} \nonumber\\
&+& \Avg{h_B(t-\Delta t)\dot{h}_B(t)h_B(t+\Delta t)}_{A_0} \nonumber\\
&+& \Avg{h_B(t-\Delta t)\dot{h}_B(t)h_A(t+\Delta t)}_{A_0}.\label{eq:Cdot_1}
\end{eqnarray}
We now condition on
the state prior to a transition, by multiplying and dividing the first two
terms by $\Avg{h_A(t-\Delta t)}_{A_0}$ and the last two terms by
$\Avg{h_B(t-\Delta t)}_{A_0}$. This gives
\begin{eqnarray}
  \dot{C}(t)&=&\Avg{h_A(t-\Delta t)}_{A_0}
  	\Avg{\dot{h}_B(t)h_A(t+\Delta t)}_{A_0,A_{t-\Delta t}} \nonumber\\
  &+&\Avg{h_A(t-\Delta t)}_{A_0}
  \Avg{\dot{h}_B(t)h_B(t+\Delta t)}_{A_0,A_{t-\Delta t}} \nonumber\\
  &+&\Avg{h_B(t-\Delta t)}_{A_0}
  	\Avg{\dot{h}_B(t)h_B(t+\Delta t)}_{A_0,B_{t -\Delta t}} \nonumber\\
  &+&\Avg{h_B(t-\Delta t)}_{A_0}
  	\Avg{\dot{h}_B(t)h_A(t+\Delta t)}_{A_0,B_{t-\Delta t}}.\label{eq:Cdot_2}
\end{eqnarray}
The average $\Avg{\dots}_{A_0,X_{t-\Delta t}}$ runs over trajectories that
start in $A$ at time $t=0$ and are in $X=A,B$ at time $t-\Delta t$.

\begin{figure}
\begin{center}
 \includegraphics[width=6cm]{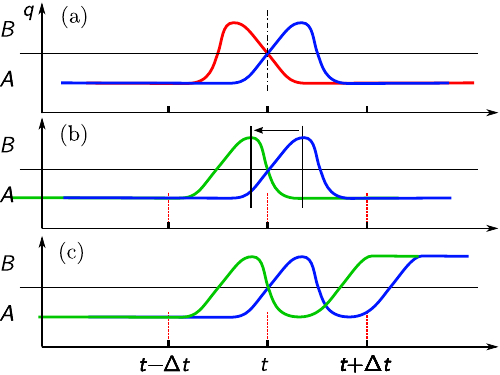}
  \caption{A visual interpretation of the term
  	$j_{AA}=\Avg{h_A(t-\Delta t)\dot{h}_B(t)h_A(t+\Delta t)}_{A_0}$ of
  	Eq.~\ref{eq:Cdot_1}.
	In microscopically reversible systems (a), for each trajectory (blue) 
	there exists an identical but time-reflected trajectory (red). 
	Consequently, $j_{AA}=0$ for all values of $\Delta t$. In non-equilibrium
	systems that are memoryless and in stationary state (b), for each 
	trajectory (blue) there exists a time-shifted trajectory (green) which 
	cancels the contribution to $j_{AA}$, so that
	$j_{AA}=0$ as well, irrespective of time-reversibility. In systems that are
	not time-reversible and remain correlated over $\Delta t$ (c), time-shifted
	trajectories may fail to contribute to $j_{AA}$, as is the case for the
	trajectory shown in green. Here cancellation cannot occur and 
	$j_{AA}\neq 0$.
	See Appendix~\ref{sec:jAA}.
\label{fig:expl}}
\end{center}
\end{figure}

Since the system looses memory on the transient time scale $\tau\ts{trans}$,
the memory fo the initial state is lost for $(t-\Delta t)>\tau\ts{trans}$,  so
that $\Avg{\dots}_{X_{t-\Delta t}, A_0}=\Avg{\dots}_{X_{t-\Delta t}}$.
In this regime Eq.~\ref{eq:Cdot_2} becomes
\begin{eqnarray}
  \dot{C}(t)&=&\Avg{h_A(t-\Delta
      t)}_{A_0}\Avg{\dot{h}_B(t)h_A(t+\Delta
      t)}_{A_{t-\Delta t}} \nonumber\\
  &+&\Avg{h_A(t-\Delta
      t)}_{A_0}\Avg{\dot{h}_B(t)h_B(t+\Delta t)}_{A_{t-\Delta
      t}} \nonumber\\
  &+&\Avg{h_B(t-\Delta
      t)}_{A_0}\Avg{\dot{h}_B(t)h_B(t+\Delta t)}_{B_{t
      -\Delta t}} \nonumber\\
  &+&\Avg{h_B(t-\Delta
      t)}_{A_0}\Avg{\dot{h}_B(t)h_A(t+\Delta
      t)}_{B_{t-\Delta t}}.\label{eq:Cdot_3}
\end{eqnarray}
We now rewrite the first and third term in this equation, which describe
trajectories that re-cross into their original state $X=A,B$ after a time
$2\Delta t$:
\begin{eqnarray}
  \dot{C}(t)&=&\frac{\Avg{h_A(t-\Delta t)}_{A_0}}{\Avg{h_A(t-\Delta t)}}
  	\Avg{h_A(t-\Delta t) \dot{h}_B(t)h_A(t+\Delta t)} \nonumber\\
  &+&\Avg{h_A(t-\Delta t)}_{A_0}
      \Avg{\dot{h}_B(t)h_B(t+\Delta t)}_{A_{t-\Delta t}} \nonumber\\
  &+&\frac{\Avg{h_B(t-\Delta t)}_{A_0}}{\Avg{h_B(t-\Delta t)}}
  		\Avg{h_B(t-\Delta t)\dot{h}_B(t)h_B(t+\Delta t)} \nonumber\\
  		&+&\Avg{h_B(t-\Delta t)}_{A_0}
  		\Avg{\dot{h}_B(t)h_A(t+\Delta t)}_{B_{t-\Delta
  		t}}\nonumber\\
  &\equiv&j_{AA}(t;\Delta t) + j_{AB}(t;\Delta t)\nonumber\\
  &+&     j_{BB}(t;\Delta t) + j_{BA}(t;\Delta t). \label{eq:Cdot_4}
\end{eqnarray}
If the system obeys microscopic reversibility, and if the initial
condition $\rho(x_0)$ coincides with the canonical distribution, then
$\Avg{\dots}$ denotes an equilibrium average. In this case the
recrossing terms 
\begin{equation}\label{eq:jAA0}
j_{AA}(t,\Delta t)=0,\;j_{BB}(t, \Delta t)=0
\end{equation}
for all times $t$ and for all values of $\Delta t$. This can be seen by noting
that the equilibrium average $\Avg{h_X(x_{t-\Delta t}) \dot{h}_B(x_t)
h_X(x_{t+\Delta t})}$ changes sign under time reversal and therefore vanishes
in thermodynamic equilibrium.  Out of equilibrium, the recrossing terms are
non-zero in general. However, under the additional condition that over the
given time scale $\Delta t$ the system is approximately time-homogeneous and
memoryless, Eq.~\ref{eq:jAA0} can be shown to hold nonetheless (Appendix
\ref{sec:jAA}). Fig.~\ref{fig:expl}
gives a graphical account of these observations.

Accepting Eq.~\ref{eq:jAA0} for the moment (this will be justified
case-by-case below), Eq.~\ref{eq:Cdot_4} reduces to
\begin{eqnarray}
  \dot{C}(t)&=&
  \Avg{h_A(t-\Delta
      t)}_{A_0}\Avg{\dot{h}_B(t)h_B(t+\Delta t)}_{A_{t-\Delta
      t}} \nonumber\\
  &+&\Avg{h_B(t-\Delta
      t)}_{A_0}\Avg{\dot{h}_B(t)h_A(t+\Delta
      t)}_{B_{t-\Delta t}} \label{eq:Cdot_5}
\end{eqnarray}

We can now relate the microscopic Eq.~\ref{eq:Cdot_5} to
the macroscopic  Eq.~\ref{eq:dPBmarkov} by identifying
$\Avg{h_B(t-\Delta t)}_{A_0} = P_B(t-\Delta t)$, 
$\Avg{h_A(x_{t-\Delta t})}_{A_0}=P_A(t-\Delta t)$ and
$\dot{C}(t)=\d P_B(t)/\d t$. 
This yields
\begin{eqnarray}
\frac{\d}{\d t}P_B (t)&=&P_A(t-\Delta t) 
	\Avg{\dot{h}_B(t)h_B(t+\Delta t)}_{A_{t-\Delta t}} \nonumber\\
&+&P_B(t-\Delta t)\Avg{\dot{h}_B(t)h_A(t+\Delta
      t)}_{B_{t-\Delta t}},\nonumber\\
&\simeq&P_A(t) \Avg{\dot{h}_B(t)h_B(t+\Delta t)}_{A_{t-\Delta t}} \nonumber\\
&+&P_B(t)\Avg{\dot{h}_B(t)h_A(t+\Delta
      t)}_{B_{t-\Delta t}}, \label{eq:dPBdt_2}
\end{eqnarray}
where the approximate equality results from the fact that $\Delta t$ is 
the time resolution of the macroscopic description.
Comparing this expression with
the macroscopic rate equation Eq.~\ref{eq:dPBmarkov}, we deduce that
\begin{eqnarray}
k_{AB}(t) &=&   \Avg{\dot{h}_B(t)h_B(t+\Delta t)}_{A_{t-\Delta
      t}},\nonumber\\
k_{BA}(t)&=&-\Avg{\dot{h}_B(t)h_A(t+\Delta
      t)}_{B_{t-\Delta t}}. \label{eq:k_AB_M}
\end{eqnarray}

In the next section, we recapitulate for completeness the classical
scenario of a Markov system that relaxes towards stationary state
starting from a non-stationary distribution \cite{chandler78}. In the
subsequent section we consider Markov systems that are driven via an
external protocol. In both cases, we will first consider the
phenomenological rate equation, and then demonstrate that the
macroscopic rate constants $k_{AB}(t),k_{BA}(t)$ are given by the
microscopic expressions on the right-hand side of Eq.~\ref{eq:k_AB_M},
provided $\Delta t$ is chosen carefully, as discussed below.

%

\subsection{Time-homogeneous Markov systems}
\label{sec:timehommarkov}

We imagine that the Markov system has been driven via some protocol into an
arbitrary initial condition $\rho(x_0)$ at $t=0$, and consider its relaxation
towards stationary state. We further imagine that experiments have revealed
that the phenomenological rate equation, Eq.~\ref{eq:dPBmarkov}, can be reduced
to
\begin{equation}
\frac{\d}{\d t}P_B(t) = k_{AB} P_A(t) - k_{BA} P_B(t),\label{eq:dPBdt}
\end{equation}
with rate constants $k_{AB}$ and $k_{BA}$; thus $k_{AB}$ is the
(constant) propensity that the system switches to $B$, given that it is in $A$.
The transitions between the states $A$ and $B$ on the macroscopic time scale
$\Delta t$ are described by a time-homogeneous Markov process.
We note that even a system with time-homogeneous
microscopic dynamics may fail to obey Eq.~\ref{eq:dPBdt}, if the system starts
from a highly non-equilibrium initial condition; indeed, the experiment has to
reveal whether the assumption of constant rates actually holds.

The solution of Eq.~\ref{eq:dPBdt} for a system that starts in state $A$
at time $t=0$ is given by
\begin{equation}
P_B (t) = P_B^{\infty} (1-e^{-t /\tau\ts{rxn}}), \label{eq:PB_t}
\end{equation}
where $\tau\ts{rxn} = (k_{AB}+k_{AB})^{-1}$ is the macroscopic
relaxation time of the switch and $P_B^{\infty}$ is the probability of
being in $B$ in the stationary state, given by $P_B^\infty =
k_{AB}/(k_{AB}+k_{BA}$).

Following the discussion above in this special case, we first need to establish
that the recrossing fluxes vanish, Eq.~\ref{eq:jAA0}. This holds whenever the
system is both memoryless and time-homogeneous for time differences larger than
$\Delta t$.
These conditions are met in the present case, because Eq.~\ref{eq:dPBdt}, which
is assumed to describe the experiment, implies that on the time scale $\Delta
t$ the system switches between $A$ and $B$ in a  memoryless fashion with
constant rates.
It then follows that the rates are given by Eq.~\ref{eq:k_AB_M}:
\begin{eqnarray}
k_{AB} &=&   \Avg{\dot{h}_B(t)h_B(t+\Delta t)}_{A_{t-\Delta
      t}},\nonumber\\
k_{BA}&=&-\Avg{\dot{h}_B(t)h_A(t+\Delta
      t)}_{B_{t-\Delta t}}. \label{eq:k_AB_1}
\end{eqnarray}
While $k_{AB}$ and $k_{BA}$ on the left-hand side are independent of
time, the expressions on the right-hand side appear to depend on $t$ and
$\Delta t$.

We first discuss the dependence on $t$. In writing down our macroscopic rate
equation, we have \emph{assumed} that the system switches between the states
$A$ and $B$ with constant rates. This presupposes that the system relaxes inside
the basins of $A$ and $B$ in between the switching events faster than the
macroscopic time $\Delta t$, leading to Markovian switching dynamics and to
loss of memory of the initial condition. Importantly, this assumption also
implies that during relaxation towards stationary state, the probability of
being in either $A$ or $B$ will change with time, but the state-space
distribution \emph{within} each macroscopic state $A$ and $B$ does not change
with time; by deduction, the state-space distribution during relaxation,
conditioned on being in either $A$ or $B$, must be equal to that of the
stationary distribution. In other words, while the ensemble brackets
$\Avg{\dots}_{A_0}$ denote an average over trajectories that start from an
arbitrary initial condition $\rho(x_0)$ in state $A$, this ensemble must,
according to our assumption, be effectively equal to the stationary ensemble of trajectories,
conditioned on starting in $A$.

This is essentially the content of Onsager's regression hypothesis
\cite{onsager31,onsager31a}, which states that the relaxation of an observable
in a non-equilibrium experiment is proportional to the  relaxation of a
spontaneous fluctuation of that observable in the equilibrium system. It relies
on the idea that the non-equilibrium initial distribution in a relaxation
experiment follows a phase-space distribution that is similar to that of a
spontaneous fluctuation in the equilibrium system \cite{chandler87}. The
regression hypothesis is a well-known theorem for systems close to thermal
equilibrium \cite{callen51}, but generalizations also exist for relaxation of
perturbations towards non-equilibrium stationary state
\cite{agarwal72,procaccia79, prost09, seifert10}. For rare switching events in
equilibrium and stationary non-equilibrium systems, the presence of a dynamical
bottleneck for switching means that the system can rapidly relax inside the
basins of $A$ and $B$, leading to loss of memory of the initial condition. This
implies that the regressions theorem also holds, with exponential
time-dependence of the relaxation function, at least on a time resolution
$\Delta t$ coarser than the transient time $\tau_{\rm trans}$.

We now turn to the dependence on $\Delta t$. Since the rates are constant, we
can time-shift Eq.~\ref{eq:k_AB_1} to give:
\begin{eqnarray}
k_{AB} &=&   \Avg{\dot{h}_B(0)h_B(\Delta t)}_{A_{-\Delta
      t}}\nonumber,\\
k_{BA}&=&-\Avg{\dot{h}_B(0)h_A(\Delta
      t)}_{B_{-\Delta t}}.\label {eq:k_AB_2}
\end{eqnarray}
These expressions, which are very similar to those of the reactive
flux method of Bennett \cite{bennett77} and Chandler \cite{chandler78}, hold
only for a certain range of $\Delta t$ values
\cite{chandler87}. The macroscopic rate constant $k_{AB}$ is defined as the
(constant) propensity that the system switches from $A$ to $B$ given that at
that moment in time it is in $A$. In contrast, the microscopic expression 
$\Avg{\dot{h}_B(0)h_B(\Delta t)}_{A_{-\Delta t}}$ is the propensity that the
system switches from $A$ to $B$ minus the propensity that it switches from $B$ 
to $A$ at a certain moment in time, given that at an earlier time $-\Delta t$
it was in $A$ and at a later time $\Delta t$ it is in $B$:
$\Avg{\dot{h}_B(0)h_B(\Delta t)}_{A_{-\Delta t}}=
\Avg{\theta[\dot{h}_B(0)]\dot{h}_B(0)h_B(\Delta t)}_{A_{-\Delta t}}
-\Avg{\theta[-\dot{h}_B(0)](-\dot{h}_B(0))h_B(\Delta t)}_{A_{-\Delta t}}$. 
Indeed, the microscopic expression takes into account that
the system may switch back and forth between the two states a number
of times, and in fact, for $\Delta t\to \infty$, 
$\Avg{\dot{h}_B(0) h_B(\Delta t)}_{A_{-\Delta t}}$ becomes equal to 
$\Avg{\dot{h}_B(\infty)}\Avg{h_B(\infty)}=0$. Clearly, the macroscopic rate
constant can only become equal to the microscopic expression if
$\Delta t$ is chosen to be smaller than the typical waiting time,
$\Delta t\ll \tau_{AB},\tau_{BA}$. 

On the other hand, $\Delta t$ cannot be made arbitrarily small. For $\Delta
t\to 0$, Eqs.~\ref{eq:k_AB_2} reduce to the transition-state approximation
\begin{eqnarray}
k_{AB}^\mathrm{TST}&=&\Avg{\theta(\dot{q})\dot{q}\delta(q-q^*)}
\nonumber,\\
k_{BA}^\mathrm{TST}&=&-\Avg{\theta(-\dot{q})\dot{q}\delta(q-q^*)}
\label{eq:k_AB_TST}.
\end{eqnarray}
Transition-state theory assumes that every trajectory that crosses the
dividing surface from $A$ to $B$ will end up in $B$ before it returns
to $A$ on a time scale $\tau_{BA}$. However, trajectories may also
recross the dividing surface a (large) number of times before they
settle in the new state $B$ or the original state $A$.  These correlated
recrossings tend to decrease the correlation function
$\Avg{\dot{h}_B(0)h_B(\Delta t)}_{A_{-\Delta t}}$ as $\Delta t$ is
increased from zero.  If the recrossings can be resolved
experimentally, and if one wishes to characterize them, then
Eq.~\ref{eq:dPBdt} is not an appropriate model and the rate constants
have to be defined differently; we will discuss
this scenario in more detail in the next section. Here, we
have been assuming that Eq.~\ref{eq:dPBdt} is an appropriate model. In this
model, the system equilibrates on time scales $\tau_A<\Delta t$ and
$\tau_B<\Delta t$ inside the states $A$ and $B$, before
switching out of these states on a much longer time scale $\tau_{AB}$
and $\tau_{BA}$, respectively.  This implies that the
correlation functions $\Avg{\dot{h}_B(0)h_B(\Delta t)}_{A_{-\Delta
    t}}$ and $\Avg{\dot{h}_B(0)h_B(\Delta t)}_{B_{-\Delta t}}$ reach a
plateau for $\Delta t$ in the range $\tau\ts{trans}<\Delta t\ll
\tau_{AB},\tau_{BA}$. 

\begin{figure}
\begin{center}
 \includegraphics[width=7cm]{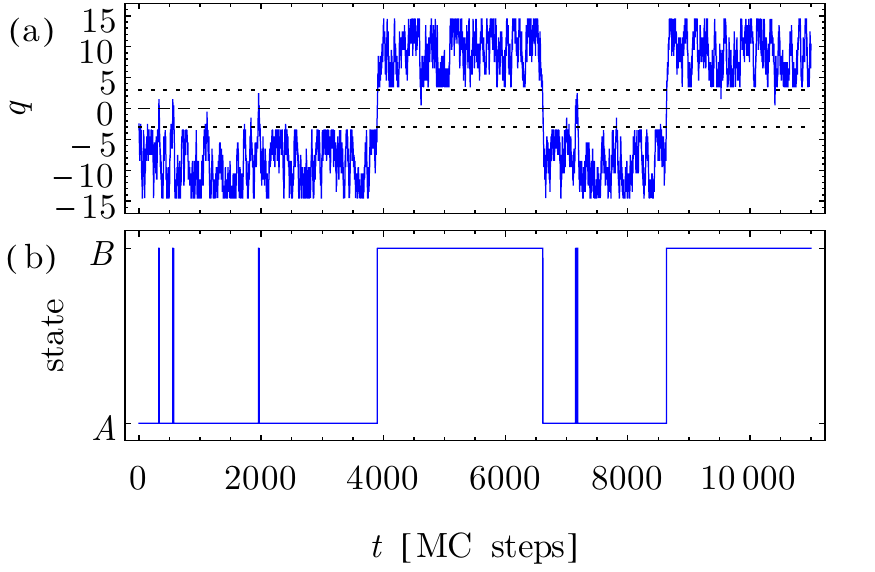}
  \caption{Microscopic (a) and macroscopic (b) trajectories in a simple
  model system of diffusion over a flat potential barrier. A particle starts
  at $q=-2.5$ and then performs a random walk visiting half-integer lattice
  points, with reflecting boundaries at $q=\pm 15$. The particle
  experiences a flat potential barrier given by $U(q)/k_BT=3\theta(2-|q|)$;
  diffusive steps are proposed in either direction and accepted according to a
  standard Metropolis criterion. The dividing surface defining the states $A$
  and $B$ is located at $q_A=q_B=q^*=0$.}
  \label{fig:trajmarkov}
\end{center}
\end{figure}

\begin{figure}[tp]
\begin{center}
 \includegraphics[width=7cm]{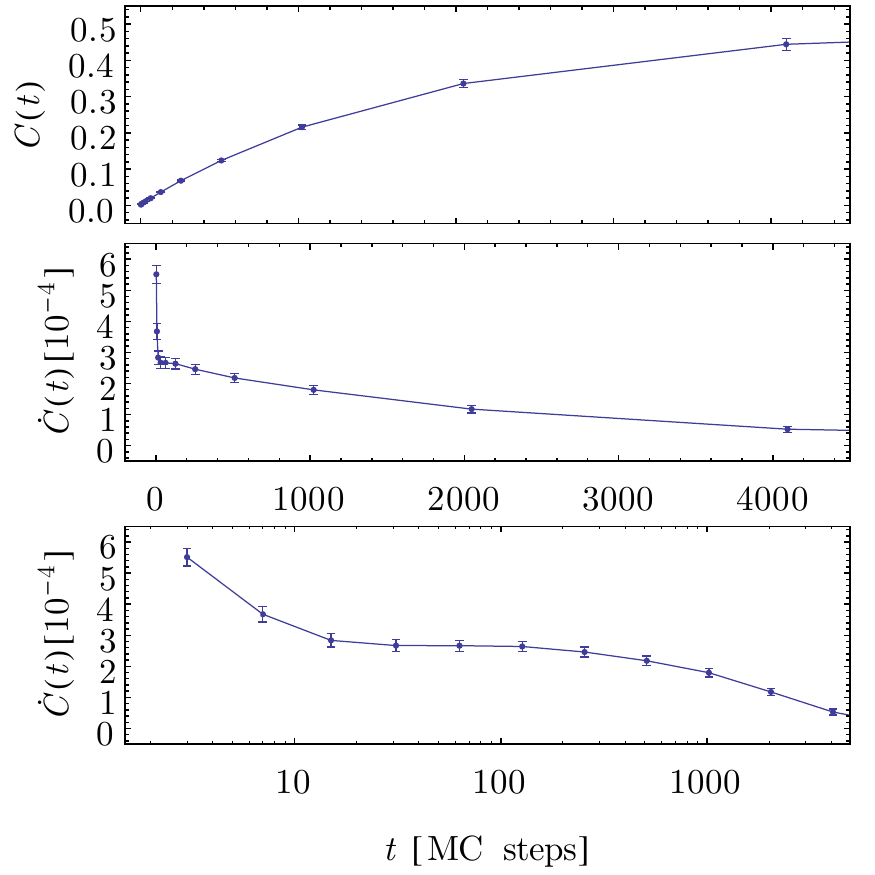}
  \caption{The correlation function $C(t)$, Eq.~\ref{eq:C_t} and its time
  derivative for the model system of Fig.~\ref{fig:trajmarkov}. Clearly on
  long times on the order of $\tau\ts{rxn}\simeq 1800$ MC steps the system
  relaxes exponentially. For very short times $t\lesssim 10$, the
  behavior is dominated by correlated transient re-crossings made by
  trajectories on the barrier top. A plateau emerges between these two times,
  whose value $\dot C\simeq 2.78\times 10^{-4}$ equals the rate constant
  $k_{AB}$.}
  \label{fig:Cfig}
\end{center}
\end{figure}

Figs.~\ref{fig:trajmarkov} and~\ref{fig:Cfig} illustrate these ideas for a
particle performing a random walk in a piecewise-flat double-well potential
according to Metropolis Monte Carlo dynamics (e.g.~\cite{frenkel01}). The
macroscopic trajectory, shown in Fig.~\ref{fig:trajmarkov}b, exhibits long
committed episodes, but also short spikes caused by transient recrossings of
the dividing surface $q^*=q_A=q_B$.
While the correlation function $C(t)=\Avg{h_B(t)}_{A_0}$ for this system shown
in Fig.~\ref{fig:Cfig}a rises exponentially for macroscopic times, its time
derivative (Fig.~\ref{fig:Cfig}b,c) does show a sudden drop for $t \lesssim
\tau\ts{trans}\approx 10$; this is due to the rapid correlated re-crossings of
the diving surface $q^*$. Importantly, for $t\ts{trans}<t< \tau\ts{rxn}$ the
derivative of the correlation function exhibits a clear plateau. In this
regime, the flux of trajectories from $A$ to $B$ is constant.
Indeed, the rate constant $k_{AB}$ is precisely given by the value of
$\dot{C}(t)$ in this regime.

In summary, we reiterate that we have not derived the macroscopic rate equation
from microscopic laws. By positing the macroscopic rate equation
Eq.~\ref{eq:dPBdt}, we have made the implicit assumption that the system
relaxes insides the basins and looses memory of its dynamics on the
experimental time scale $\Delta t$. The same assumption also entails that the
non-stationary ensemble of trajectories is similar to the stationary ensemble
of trajectories, conditioned on starting in $A$. In other words, if the
macroscopic rate equation gives an accurate description of the switching
dynamics, then this strongly suggests that the Onsager regression theorem
holds as well.

\subsection{Externally driven Markov systems}

We now consider time-inhomogeneous Markov systems. We imagine that from $t=0$
the Markovian system is under the influence of some time-dependent external
force $\phi(t)$ that biases it towards one state or the other. The macroscopic
rate equation thus has the general form of Eq.~\ref{eq:dPBmarkov}, and is valid
for times longer than a macroscopic time $\Delta t$ as inferred from
experiment. The general solution can be written as
\begin{equation}\label{eq:PABtimeinmarkov}
P_B(t)=P_B(0) e^{-\int_0^t \frac{ \d t'}{\tau\ts{rxn}(t')}}+ \int_0^t
k_{{AB}}(t') e^{-\int_{t'}^t \frac{\d t''}{\tau\ts{rxn}(t'')} } \d t',
\end{equation}
where we define $\tau\ts{rxn}^{-1}(t)=k_{AB}(t)+k_{BA}(t)$.

Examples of externally driven Markov systems include protein unfolding
driven by a time-dependent force and (crystal) nucleation at a
time-varying temperature. This class also contains systems where one
(or more) of the degrees of freedom relaxes slowly on the time scale
of a switching event, such as gene networks where at a given time a
gene is turned on, which then induces the flipping of a genetic
switch. The protein that drives the flipping of the switch could then
be viewed as the external force $\phi(t)$ that acts on the switch. We
consider external protocols with finite band-width such that there
exist a scale of fastest force variation $\tau_\phi$ and a scale of
slowest variation $T_\phi$, where $T_\phi$ is at most the duration of the
experiment, $T_\phi\leq T$.

What microscopic scenarios are compatible with the macroscopic rate law,
Eq.~\ref{eq:dPBmarkov}? The rate law implies that the system looses microscopic
memory beyond $\Delta t$, so that $\tau\ts{trans}<\Delta t$ must hold. As
before, we restrict our attention to systems with a time scale separation
between intrastate relaxation and global relaxation, $\tau\ts{trans}\ll
\tau_{AB,BA}$, where $\tau_{AB}$ and $\tau_{BA}$ are the mean waiting times for
switching  \footnote{Since the transition rates are now
  time-dependent, the definition of the waiting times for transitions,
  $\tau_{AB}$ and $\tau_{BA}$ is less clear-cut. A way to obtain these time
  scales is to take the average `instantaneous waiting times' over the duration
  of the protocol.
  E.g.~$\tau_{AB} = 1/T\int_0^T k_{AB}(t)^{-1}\d t$.}.
  Under these constraints, there still is a variety of
possible relations between the time scales of driving and the intrinsic
microscopic time scales of the system.

The simplest situation is the quasi-static case: While the
state-space distribution evolves with $\phi(t)$ in between the switching
events, the system can locally (i.e. within the macroscopic states) adapt to
the force, meaning that
$\tau\ts{trans}<\tau_\phi$. Importantly, this quasi-static forcing scenario
does not require the force to be slower than the global relaxation;
$\tau_\phi,T_\phi \gtrless \tau_{AB},\tau_{BA}$. We then expect that it
should be straightforward to define time-dependent rate functions in terms of 
microscopic correlation functions.

To derive the microscopic expressions that are consistent with the macroscopic
Eq.~\ref{eq:dPBdt}, we first note that in this quasi-static case we may choose
$\Delta t$ such that $\tau\ts{trans} < \Delta t < \tau_\phi,\tau\ts{rxn}$. 
On this time scale $\Delta t$ the macroscopic
switching dynamics is memoryless and approximately time-homogeneous, since the
system adapts instantaneously to the force. Thus Eq.~\ref{eq:jAA0} applies and
the expressions Eqs.~\ref{eq:Cdot_3}-\ref{eq:dPBdt_2} hold unchanged.
We then indeed arrive at Eq.~\ref{eq:k_AB_M}. The rates now depend on time $t$,
but since the force is quasi-static, the dynamics in each $\Delta t$ interval
is governed only by the value of $\phi$. Time dependence thus enters only via
$\phi(t)$: 
\begin{eqnarray}
k_{X\oX}(t)&=&k_{X\oX}(\phi(t))
\text{, if }\tau\ts{trans}< \Delta t < \tau_\phi.\label{eq:viaphi}
\end{eqnarray}
Furthermore, when $\Delta t$ is in the range $\tau\ts{trans}<\Delta t
<\tau_\phi,\tau\ts{rxn}$, we can expect that the correlation functions in
Eq.~\ref{eq:k_AB_M} are constant as a function of $\Delta t$, in analogy to
the time-homogeneous case.

\begin{figure}[htp]
\begin{center}
 \includegraphics[width=7cm]{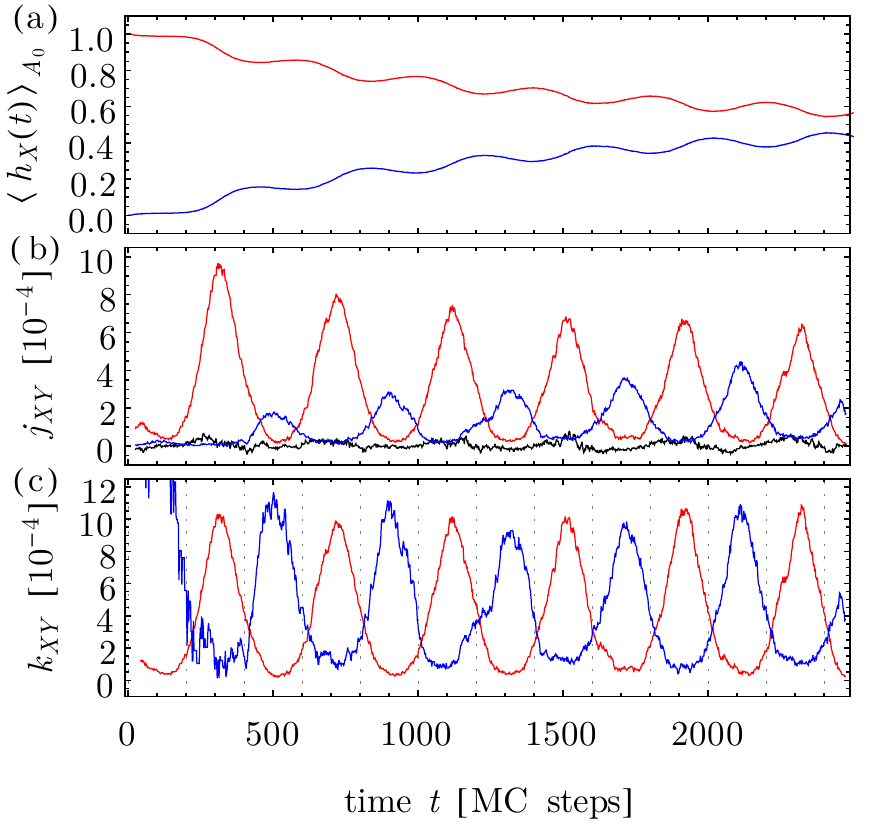}
  \caption{Time-dependent version of the system of
    Fig.~\ref{fig:trajmarkov}, under the influence of an external
    force. The system is started in stationary state within $A$ and
    relaxes with a time dependent modulation (a; $\Avg{h_A}$, blue,
    $\Avg{h_B}$, red). Fluxes defined in Eq.~\ref{eq:Cdot_1} (b;
    $j_{AB}$, blue, $j_{BA}$, red, $j_{AA}$, black) reflect the state
    occupancies.  The rate functions, Eq.~\ref{eq:k_AB_M} (c; colors
    as in b) are independent of state occupancy. At the
    zeros of the force (dotted lines) the time dependent rates
    coincide with the rate constant of the stationary system,
    $k_{AB}=2.78\times10^{-4}$ (see Fig.~\ref{fig:Cfig}), in agreement with 
    a quasi-stationary description on the time scale of the driving.}
  \label{fig:jABforced}
\end{center}
\end{figure}

Fig.~\ref{fig:jABforced} illustrates these relations in a version of the
barrier-diffusion toy model, which is now augmented with an oscillatory force
with a single frequency, $f(t) = a k_BT \sin(2\pi t/\tau_\phi)$ with
$\tau_\phi=T_\phi=400$, $a=0.1$. We choose $\Delta t=20$ so that
$\tau\ts{trans}<\Delta t\ll \tau_\phi,\tau\ts{rxn}$. This separation of time
scales suggests that we can treat this time-inhomogeneous Markov system as
quasi-stationary, so that  $j_{AA}\simeq 0$ (Eq.~\ref{eq:jAA0}), and time
enters the rates only via the force. Indeed, we observe that the recrossing
fluxes approximately vanish, and that at the times when the force crosses zero,
the rate constant equals that of the time-homogeneous Markov system of
Fig.~\ref{fig:Cfig}, supporting this idea.

An alternative simple scenario arises in the opposite case of rapid driving,
where the external force variations are faster than the local relaxation within
the macroscopic states, i.e.~$\tau_\phi<T_\phi<\tau\ts{trans}$. We can then
choose a macroscopic resolution $\Delta t > \tau_\phi,
T_\phi,\tau\ts{trans}$, so that the macroscopic description effectively
averages both over microscopic correlations of the system and over the
variations in driving. In this case the macroscopic switching dynamics is
described by the phenomenological rate equation of the time-homogeneous system,
Eq.~\ref{eq:dPBdt}, although the magnitudes of the rate constants, which are
independent of $\Delta t$, will be renormalized by the time-varying force.
Moreover, because the system is effectively time-homogeneous and switches in a
memoryless fashion between $A$ and $B$ , again Eq.~\ref{eq:jAA0} holds, so that
the rate constants are given by the plateau values of the correlation functions
in Eqs.~\ref{eq:k_AB_2} in the regime $\tau_\phi,\tau\ts{trans}<\Delta t\ll
\tau_{AB},\tau_{BA}$.

In the previous two scenarios we were able to choose $\Delta t$ to be both
well-separated from the driving time scales and within a well-defined plateau
of the derivative of the microscopic correlation function; this allowed us to
extract the rate as $k_{AB}=\dot C(\Delta t)$, irrespective of the precise
value of $\Delta t$. We now consider a scenario in which this is not
possible. Suppose that a certain macroscopic time resolution $\Delta t$ is
required, which still leads to Markovian switching ($\tau\ts{trans}<\Delta t$)
but lies \emph{within} the frequency band of the driving, $\tau_\phi < \Delta
t<T_\phi$. For instance, we may wish to describe a non-stationary experiment
carried out with a given time resolution equal to $\Delta t$. 

In the case that the microscopic relaxation of the system is the fastest
process, $\tau\ts{trans} <\tau_\phi$ as in the quasi-stationary case studied
above, then the correlation function does exhibit a plateau, but in a regime
$\tau\ts{trans} <\widetilde{\Delta t} < \tau_\phi$ below the macroscopic
resolution $\Delta t$. While this does allow us to uniquely define a
force-dependent rate constant as $k_{AB}(\phi)=\dot{C}(\widetilde{\Delta t})$,
this rate constant is inaccessible in experiments on the time scale $\Delta t$.
The macroscopic rate constant as measured in the experiment now depends on the
experimental time resolution $\Delta t$:
\begin{equation}
k_{AB}^{\rm eff}(t)\simeq (\Delta t)^{-1}\int_{t-\Delta t/2}^{t+\Delta t/2}
k_{AB}(\phi(t'))\d t'.\label{eq:kABeffmixed1}
\end{equation}

Alternatively, when the driving contains components that are faster than
microscopic relaxation, $\tau_\phi<\tau\ts{trans}$, the system is Markovian on
the time scale $\Delta t$ but retains memory of the driving over times
$\tau\ts{trans}$. In this case ($\tau_\phi<\tau\ts{trans}<\Delta t<T_\phi$) the
correlation function $C(t)$ does not exhibit a well-defined plateau on any time
scale, and, as a result, a rate constant cannot be defined in a unique,
protocol-independent manner. It is of course still possible to define a rate
function in terms of a first passage time density, as a function of the
external protocol; for analytical treatments in model
systems with simple driving protocols see for instance \cite{lindner04} and
references therein.

\section{Non-Markovian systems} \label{sec:NonStatSys}

For equilibrium systems, the two observations of rapid switching
$\tau_C\ll\tau_{AB},\tau_{BA}$ (Eq.~\ref{eq:tau_C}) and rapid equilibration
$\tau_A\ll\tau_{AB};\tau_B\ll\tau_{BA}$ (Eq.~\ref{eq:tau_A_B}) are often
intimately connected. Rare events typically arise because of a single large
free-energy barrier separating the two states $A$ and $B$. As a consequence,
the waiting time tends to be much longer than the switching event
itself---making a two-state description meaningful---while it also allows the
system to relax inside the basins of $A$ and $B$ in between the switching
events---leading to memoryless switching on the macroscopic time scale $\Delta
t$. In these equilibrium systems there is only one relevant macroscopic time
scale $\tau\ts{rxn}$, which is associated with the global relaxation of state
occupancies. However, even for equilibrium systems, inertial effects may lead
to correlated recrossings \cite{erp11}, possibly giving rise to non-Markovian
switching dynamics on the macroscopic time scale $\Delta t$; if such correlated
recrossings are important, then the macroscopic model of Eq.~\ref{eq:dPBdt}
needs to be refined.

One can readily find examples of non-equilibrium
systems, however, that can be coarse grained as two-state systems, but in
which switching between the two states occurs in a non-Markovian fashion 
even on time scales that are comparable to
the typical waiting times $\tau_{AB}, \tau_{BA}$. This means that while the
waiting time is much longer than the duration of the switching event, there
exists another time scale in the system that is comparable to the waiting time
for switching.

An example of such a non-Markovian, non-equilibrium system is the bacterial
flagellar motor.  This rotary motor can run in either clockwise or
counterclockwise direction.  The transitions between these two states
happen much faster than the typical waiting time for switching,
suggesting a two-state description. In experiments, the distribution
of waiting times was observed to be non-exponential, and the power
spectrum of the switching dynamics featured a peak around $1 \mathrm
s^{-1}$~~\cite{korobkova06}, which means that there is a characteristic
frequency at which the motor switches.

These data show that the flagellar motor cannot be described as a random
telegraph process in which the switching events are independent, and the
clockwise and counterclockwise intervals are uncorrelated and exponentially
distributed; indeed, this system cannot be described by Eq.~\ref{eq:dPBdt}. The
data also imply that motor switching is coupled to a non-equilibrium process
\cite{kampen92,tu05,albada09}. It appears that the conformational dynamics of
the rotor proteins is coupled to the slow relaxation dynamics of the flagellum,
which in turn is influenced by the rotation of the motor, driven by a proton
motive force \cite{albada09}.

To describe systems like the bacterial flagellar motor in which memory is
important, we now consider non-Markovian systems that when left unperturbed,
reach a stationary state (equilibrium or not). As before, we imagine that the
system flips between two macroscopic states $A$ and $B$ that are metastable in
the sense that the waiting time between the switching events is much longer
than the duration of the switching event itself. These system have memory on
the macroscopic time scale $\Delta t$ which means that the propensity of
switching between the two macroscopic states at a given moment in time depends
upon the history of the switching dynamics.

To derive microscopic expressions for the rate functions in the presence of
memory, we start again with the phenomenological description of the system. If
the propensity to switch depends upon the macroscopic history even prior to
the last switching event, then Eq.~\ref{eq:ME} cannot be simplified much
further, and remains of little use. However, it is conceivable that the
switching propensity depends upon the macroscopic history only since the last
switching event, to a good approximation. The bacterial flagellar
motor is an example of such a system: when
the experimental clockwise and counterclockwise intervals of the flagellar
motor were randomly shuffled, the power spectrum was unchanged; moreover, the
 observed power spectrum could be reproduced from the measured
waiting-time distributions only \cite{korobkova06}. This strongly suggests that
the different intervals are in fact temporally uncorrelated, so that the
switching propensity only depends upon the time that has passed since the last
switching event. We refer to this as the \emph{clock-resetting}
scenario---after each switching event, the system looses memory; in other
words, switching is a renewal process. In what follows we will restrict our
attention to this case. We do not exclude at this point the presence of
external driving; in that case the system has only a single-switch memory but
is time-inhomogeneous.

In the clock-resetting scenario, the phenomenological rate equation 
Eq.~\ref{eq:ME} can be simplified:
\begin{equation}
\frac{\d}{\d t}P_B(t)=\int_{t>t'}k_{A
  B}(t|t')P_A(t;t') - k_{BA}(t|t')P_B(t;t')\d t'.
\label{eq:dPBdt_NM}
\end{equation}
Here, $P_A(t;t')\d t'$ is the joint probability that the system is in
state $A$ at time $t$ and has switched into this state for the last
time within the earlier interval $(t',t'+\d t')$. It can be decomposed
as the propensity $q_{BA}(t')$ to enter $A$ at $t'$,  times
the survival probability $S_A(t|t')$ to stay in $A$ until at least time $t$:
$P_A(t;t')=q_{BA}(t')S_A(t|t')$. 

The rate kernel $k_{AB}(t|t')$ is defined as the propensity
that the system switches from $A$ to $B$ at time $t$
given that it has switched into $A$ at time $t'<t$ and is
still in $A$ at $t$. It is given
by the propensity $q_{AB}(t|t')\equiv -\partial_t S_A(t|t')$ that a trajectory
that entered $A$ at time $t'$, switches from $A$ to $B$ at a later time $t$,
divided by the probability that it is still in $A$ at time $t$:
$k_{AB}(t|t')=q_{AB}(t|t')/S_A(t|t')=-\partial_t\ln S_A(t|t')$. If the
system is time-homogeneous, i.e.~without external time dependence,
then we have $k_{AB}(t|t')=k_{AB}(t-t')$. If the system is in
stationary state, then furthermore $P_B(t;t')=P_B(t-t')$. We note that
this phenomenological model can be solved analytically
\cite{albada09}, see also Appendix \ref{sec:nonmarkovmacrosol}.

Before we derive microscopic expressions for the rate kernels 
that are consistent with Eq.~\ref{eq:dPBdt_NM}, we first discuss the
time scales that are relevant in such a system. A two-state
description implies that the duration of the switching event is much
shorter that the waiting time for switching:
$\tau_C\ll\tau_{AB},\tau_{BA}$. Moreover, we expect that inside the
states $A$ and $B$ most degrees of freedom relax quickly, within
$\tau_A$ and $\tau_B$ respectively, where $\tau_A\ll\tau_{AB}$ and 
$\tau_B\ll\tau_{BA}$; we can thus identify a fast transient time scale
$\tau\ts{trans} \simeq \max\{\tau_C,\tau_A,\tau_B\}$. If these were
the only time scales in problem, then the macroscopic switching
dynamics could be described as that of a two-state Poisson process,
assuming that we choose a macroscopic time scale $\Delta
t>\tau\ts{trans}$.  However, the observation that the macroscopic
switching dynamics exhibits memory on the scale $\Delta t$, implies
that there is another time scale in the problem, $\tau\ts{slow}$,
which is at least on the order $\tau\ts{slow}\gtrsim \Delta t$.

The bacterial flagellar motor provides a concrete illustration of these ideas:
while the switching dynamics of the motor in the presence of the flagellum is
non-Poissonian, the dynamics in the absence of the flagellum is Poissonian,
with constant rates, leading to exponential waiting-time distributions
\cite{bai10}. The reason for this change in behavior is that the relaxation
dynamics of the flagellum introduces a slow time scale, which
becomes comparable to, and can even set, the typical waiting time for
switching. After a switching event, the flagellum first unwinds, but then,
driven by the rotation of the motor, winds up in the new direction; this leads
to an increase in the force on the rotor proteins, which then tends to switch
their conformation, and thereby the rotation direction of the motor
\cite{albada09}.

\begin{figure}[htp]
\begin{center}
 \includegraphics[width=7.5cm]{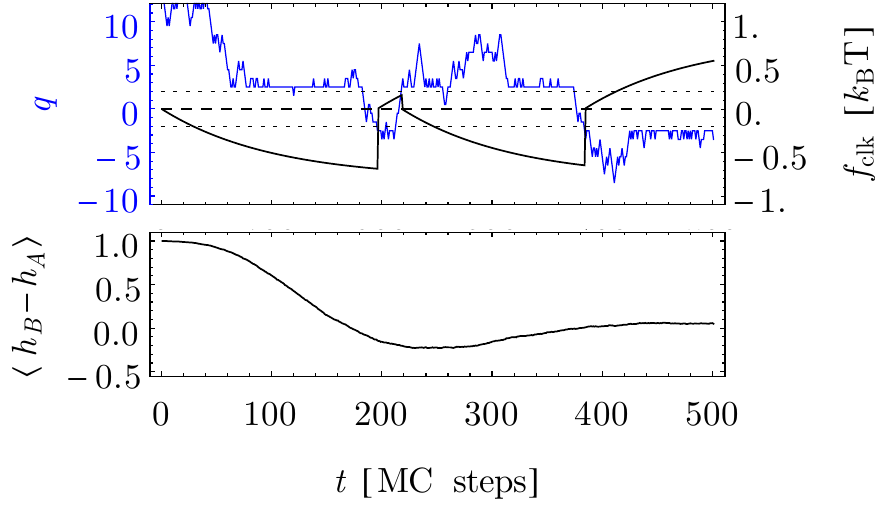}
  \caption{Particle diffusing with residence-time dependent restoring force.
  A trajectory (top, blue, left axis) reflects the restoring force (black,
  right axis). The relaxation of state occupancies when started uniformly in
  the $B$ state basin (bottom) is oscillatory, and more rapid than without
  force (cf.~Fig.~\ref{fig:Cfig}).}
  \label{fig:forcetraj}
\end{center}
\end{figure}

We extend our diffusive-barrier crossing model to capture the non-equilibrium
switching of the bacterial flagellar motor, see Fig.~\ref{fig:forcetraj}. The
particle diffuses in a piecewise constant potential with a single flat barrier
of height $4k_\mathrm{B}T$ at $|q|<2$; the state boundaries are at
$q_A=q_B=q^*=0$. Whenever the particle enters one of the two potential wells
(say, at $t'$), the clock is reset and a restoring force ramps up over time,
$f\ts{clk}=-\mathrm{sgn}(q) f_0 [1-e^{-(t-t')/\tau_\phi}]$ with time constant
$\tau_\phi=100$ and saturating magnitude $f_0=0.8k_\mathrm{B}T$.
The addition of a restoring force leads to a faster global relaxation, which
for the chosen value of $f_0$ is on the order of $\tau_\phi$.
Fig.~\ref{fig:forcetraj} shows that the relaxation of this system is not
exponential: There is an initial lag time, and relaxation occurs in an
oscillatory fashion, because the restoring force introduces a characteristic
time scale for switching, and because particle clocks are taken to
be synchronized at $t=0$.
We remark that an exponential relaxation observed in a single experiment would
\emph{not} by itself imply memoryless switching: For instance, the same system
with unsynchronized clocks at $t=0$ relaxes exponentially (not shown).

To derive microscopic expressions for the rate kernels that are
consistent with the macroscopic rate equation of Eq.~\ref{eq:dPBdt_NM}, we need
to define an indicator function that serves to measure the time since the last
switching event. We let $H_X(t,t')\equiv\prod_{t\geq t'' > t'}h_X(t'')$. By
construction, $H_X(t,t')$ takes the value 1 if the given trajectory was in
$X=A,B$ without interruption from $t'$ to $t$ and 0 otherwise, and $H_X(t,t) =
h_X(t)$. One also verifies (Appendix \ref{sec:Hmath}) that
\begin{eqnarray}
\partial_t H_X(t,t') &=& \dot{h}_X(t)
  H_X(t,t') \text{, and}\nonumber\\
\partial_{t'} H_X(t,t')&=&
  H_X(t,t')\dot{h}_X(t').\label{eq:dot_H_X}
\end{eqnarray}
The temporal derivatives of $H_X(t,t')$ with respect to $t'$ and $t$
(Eq.~\ref{eq:dot_H_X}) thus make it possible to count the last entry time and
the first exit time out of $X$, respectively. Although the definition of $H$
may appear convoluted, it is straightforward to measure in a simulation, by
recording switching times along trajectories. Interestingly an
indicator function similar to $H$ has been employed by Jung \textit{et al} 
\cite{jung05} to characterize fluctuations of trajectories in
glassy systems.

Applying the relations of Eq. \ref{eq:dot_H_X} and the identity
$\dot{h}_A+\dot{h}_B=0$, we can obtain the following relation:
\begin{subequations}\label{eq:totalhBdot}
\begin{eqnarray}
\dot{h}_B(t)&=&\dot{h}_B(t)\sum_{X=A,B}H_X(t,t)\nonumber\label{eq:dot_h_B}\\
&=&\dot{h}_B(t)\Bigl[\sum_{X=A,B} H_X(t,t_0) + \int_{t_0}^t \partial_{t'}
H_X(t,t')\d t'\Bigr]\nonumber\\
&=&\partial_t H_B(t,t_0) - \partial_t H_A(t,t_0) \label{eq:H_X_0}\\
&&+\int_{t_0}^t\partial_t\partial_{t'} H_B(t,t')
  - \partial_t\partial_{t'} H_A(t,t') \d t'.\label{eq:H_AB_t}
\end{eqnarray}
\end{subequations}
Here the probability flux in and out of $B$ is partitioned into contributions
from trajectories that remained in one state since $t_0$ and then switch at $t$
(Eq.~\ref{eq:H_X_0}), and trajectories that had their last switching event at
$t'>t_0$ before switching at $t$ (Eq.~\ref{eq:H_AB_t}). The latter terms allow
us to derive correlation functions which express the rate kernels 
depending on the last switching event.


To derive a microscopic expression that is consistent with the
 the macroscopic equation \ref{eq:dPBdt_NM}, we take
the ensemble average $\Avg{\dots}$ of Eq.~\ref{eq:totalhBdot} (see also
Eq.~\ref{eq:Cdot_0}). The ensemble average is defined by the initial
distribution $\rho(x_0)$ as before (although here we do not condition on
starting in state $A$), and possibly by a prescription for $t<0$, as discussed
below. We now make the identifications
\begin{subequations}\label{eq:nonMarkovidentify}
\begin{eqnarray}
\Avg{h_B(t)}&=&P_B(t),\\
\partial_{t'}\Avg{H_X(t,t')}\d t'
&=&\Avg{H_X(t,t')\dot{h}_X(t')}\d t' \nonumber\\
&=&P_X(t;t')\d t',\\
\frac{\partial_t\partial_{t'}\Avg{H_X(t,t')}}
{\partial_{t'}\Avg{H_X(t,t')}} &=&
\frac{\Avg{\dot{h}_X(t)H_X(t,t')\dot{h}_X(t')}}
{\Avg{H_X(t,t')\dot{h}_X(t')}}\nonumber\\
&=&-k_{X\oX}(t|t'), \label{eq:kdefstrict}
\end{eqnarray}
\end{subequations}
where $X=A,B$ and $\oX=B,A$. We can then rewrite the microscopic equation
\ref{eq:totalhBdot} in the form of Eq.~\ref{eq:dPBdt_NM}.
\begin{subequations}\label{eq:dPBdt_NM_BC}
\begin{gather}
\frac{\d }{\d t}P_B(t)=
\partial_t\Avg{H_B(t,t_0)}-\partial_t\Avg{H_A(t,t_0)}
\label{eq:dPBdt_NM_BC_bdry}\\
+\int_{t_0}^t k_{AB}(t|t')P_A(t;t')-k_{BA}(t|t')P_B(t;t')\d t'.
\label{eq:dPBdt_NM_BC_int}
\end{gather}
\end{subequations}
The additional boundary terms Eq.~\ref{eq:dPBdt_NM_BC_bdry} appear here due to
a finite lower integration limit $t_0$.  They can be treated in two ways.  The
first option is taking $t_0\to-\infty$. Then the terms 
Eq.~\ref{eq:dPBdt_NM_BC_bdry} vanish, because the system is not expected to
stay in a given state forever. To extract rate constants from a
microscopic measurement, we then have to be able to determine the integrand in
Eq.~\ref{eq:dPBdt_NM_BC_int} for negative $t'$. This will be possible for
instance if we are dealing with a time-homogeneous system which was prepared
in a stationary state for negative times, since then $P_X(t;t')=q_{\oX X,
\mathrm{ss}}S_X(t-t')$ for all $t'<0$, with a constant, stationary-state
value of the influx $q_{\oX X, \mathrm{ss}}$; see also Appendix
\ref{sec:nonmarkovmacrosol}.

Instead, for non-stationary systems one may wish to explicitly account for the
switching events only after $t>0$ without specifying the history before $t<0$.
Then one would let $t_0\to0$ and retain the boundary terms,
Eq.~\ref{eq:dPBdt_NM_BC_bdry}. They give a transient contribution to the
macroscopic evolution of the system, which summarizes how the system was
prepared before $t=0$. We can incorporate this contribution by defining
\begin{subequations}
\begin{eqnarray}
\frac{\partial_t
  \Avg{H_X(t,0)}}{\Avg{H_X(t,0)}}&=&-k_{X\oX}^{<0}(t),\\
\Avg{H_X(t,0)}&=&P_X^{<0}(t).
\end{eqnarray}
\end{subequations}
In this case, Eq.~\ref{eq:dPBdt_NM_BC} becomes
\begin{subequations}\label{eq:dPBmarkovnonmarkov}
\begin{gather}
\frac{\d }{\d t}P_B(t)=k_{AB}^{<0}(t)P_A^{<0}(t)-k_{BA}^{<0}(t)P_B^{<0}(t)
\label{eq:dPBmnmm}\\
+\int_0^tk_{AB}(t|t')P_A(t;t')-k_{BA}(t|t')P_B(t;t')\d t',\label{eq:dPBmnmnm}
\end{gather}
\end{subequations}
which has an effective Markovian part (Eq.~\ref{eq:dPBmnmm}) and a
non-Markovian (Eq.~\ref{eq:dPBmnmnm}) part.


Scrutinizing the arguments above more carefully, we should expect that the
system exhibits transient recrossings of the dividing surface $q^*$ that are
not persistent on the macroscopic time scale $\Delta t$. Moreover, these
transient recrossings are not expected to lead to loss of memory: only
macroscopic switching events, in which the systems fully transitions from one
basin of attraction to another, are likely to reset the clock. This means that
the identifications made in Eqs.~\ref{eq:nonMarkovidentify}, like the TST
expressions Eqs.~\ref{eq:k_AB_TST}, are erroneous in that they overcount
transient switching events. An additional point concerns the experimental time
resolution. It is conceivable that the experimental resolution is sufficiently
high that transient crossing events can be detected that do not reset the
clock. If one wishes to describe these events, then the phenomenological model
of Eq.~\ref{eq:dPBdt_NM}, which only describes the crossing events that lead to
loss of memory, needs to be modified. Here, we consider the case that
Eq.~\ref{eq:dPBdt_NM} is an appropriate macroscopic model. The task is then to
come up with a microscopic description that integrates over the transient
recrossings of the dividing surface that do not reset the clock.


To integrate over the transient recrossings of the diving surface that
separates the macroscopic states $A$ and $B$, we must modify the indicator
function $H_{X\oX}$.  This can be achieved by introducing a  `grace interval'
$\Delta t$ over which such short excursions are tolerated. Therefor we define a
relaxed version of $h_X$ as
\begin{equation}\label{eq:h_Xtdt}
h_X(t;\Delta t) = \theta\bigl[\int_{t-\Delta t}^th_X(t')\d t'-\Delta t/2\bigr];
\end{equation}
this new indicator function takes the value 1 precisely if in the preceding
short interval, the system was predominantly in state $X$; the arbitrary
convention of using preceding and not e.g.~centered intervals becomes
unimportant on the macroscopic scale.
Observe that, as for $h_X$, we have $h_A(t;\Delta t)+h_B(t;\Delta t)\equiv 1$
(we still take the states to be adjacent, $q_A=q_B=q^*$); the previous
definition $h_X(t)$ is recovered in the limit $\Delta t\to 0$.
Using this coarse-grained state indicator function, the indicator for
uninterrupted residence in a state is then defined as before,
\begin{equation}
H_X(t,t';\Delta t)=\prod_{t \geq  t'' > t'}h_{X}(t'';\Delta t).
\end{equation}
This expression remains 1 for instance if all excursions from $X$ between
$t'-\Delta t$ and $t$ are shorter than $\Delta t/2$. By definition,
$H_X(t,t';\Delta t)$ is a monotonically decreasing function of $t$.
Furthermore, $H_X(t,t;\Delta t)=h_X(t;\Delta t)$, and $H_X(t,t';\Delta t) \to
H_X(t,t')$ as $\Delta t\to 0$.

\begin{figure}[htp]
\begin{center}
 \includegraphics[width=7cm]{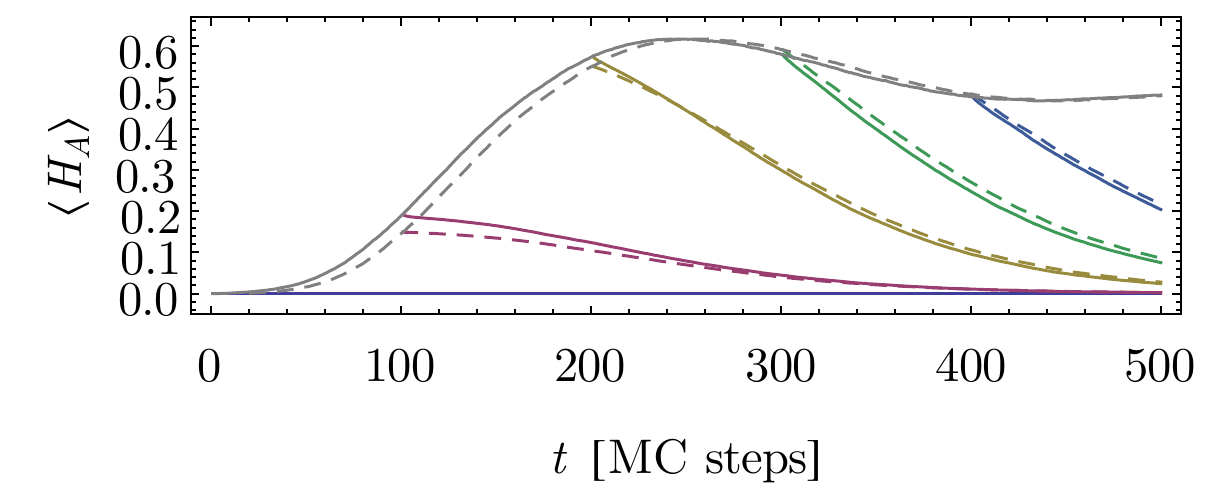}
  \caption{Uninterrupted state occupancies $\Avg{H_A(t,t';\Delta t)}$ for the
  system from Fig.~\ref{fig:forcetraj}, for $\Delta t=0,20$ (solid and dashed
  lines, respectively). The system starts in state $B$ and 
  accumulates in $A$ with occupancy $\Avg{H_A(t,t;\Delta
  t)}-\Avg{h_A(t;\Delta t)}$, shown in gray.
  Crossings back to $B$ then decrease $\Avg{H_A(t,t';\Delta t)}$, shown in
  colors for fixed values $t'=0,100,200,400$ from left to right.}
  \label{fig:HXplot}
\end{center}
\end{figure}

Fig.~\ref{fig:HXplot} shows the average uninterrupted state occupancies for the
model system introduced above. While the state occupancy $H_A(t,t;\Delta
t)=h_A(t;\Delta t)$ displays the oscillatory global relaxation in this system,
$H_X(t,t';\Delta t)$ decays with increasing residence time $t-t'$. This decay
is slower when $\Delta t$ is finite due to the fact that brief excursions from
the state are tolerated.

With the new definitions for $h$ and $H$, the equalities in
Eq.~\ref{eq:totalhBdot} still hold. Therefore, also Eq.~\ref{eq:dPBdt_NM_BC}
holds unchanged if we make the new identifications
\begin{subequations}\label{eq:nonMarkovidentifydelta}
\begin{eqnarray}
\Avg{h_B(t, \Delta t)} &=& P_B(t), \\
\partial_{t'} \Avg{H_X(t,t';\Delta t)}\d t' &=&P_X(t;t';\Delta t)\d t',\\
\frac{\partial_t \partial_{t'}\Avg{H_X(t,t';\Delta t)}}
{\partial_{t'}\Avg{H_X(t,t';\Delta t)}}&=&-k_{X\oX}(t|t';\Delta t),
\label{eq:kdefgrace}
\end{eqnarray}
\end{subequations}
replacing Eqs.~\ref{eq:nonMarkovidentify}.

If the system exhibits transient recrossings on a time scale $\tau\ts{trans}\ll
\tau\ts{slow}$,  then $k_{X\oX}(t|t';\Delta t)$ should become independent of
$\Delta t$ if we choose $\Delta t$ to be in the range $\tau\ts{trans}<\Delta
t<\tau\ts{slow}$. Moreover, we can expect that with this choice of $\Delta t$,
the rate constant $k_{X\oX}(t|t';\Delta t)$ agrees with the experimentally
measured rate constant $k_{X\oX}(t|t')$ when the experimental time
resolution is indeed of the order $\Delta t$, such that the transient
recrossings are not detected, while the presence of a slow time scale is
detected.

\begin{figure}[htp]
\begin{center}
 \includegraphics[width=7cm]{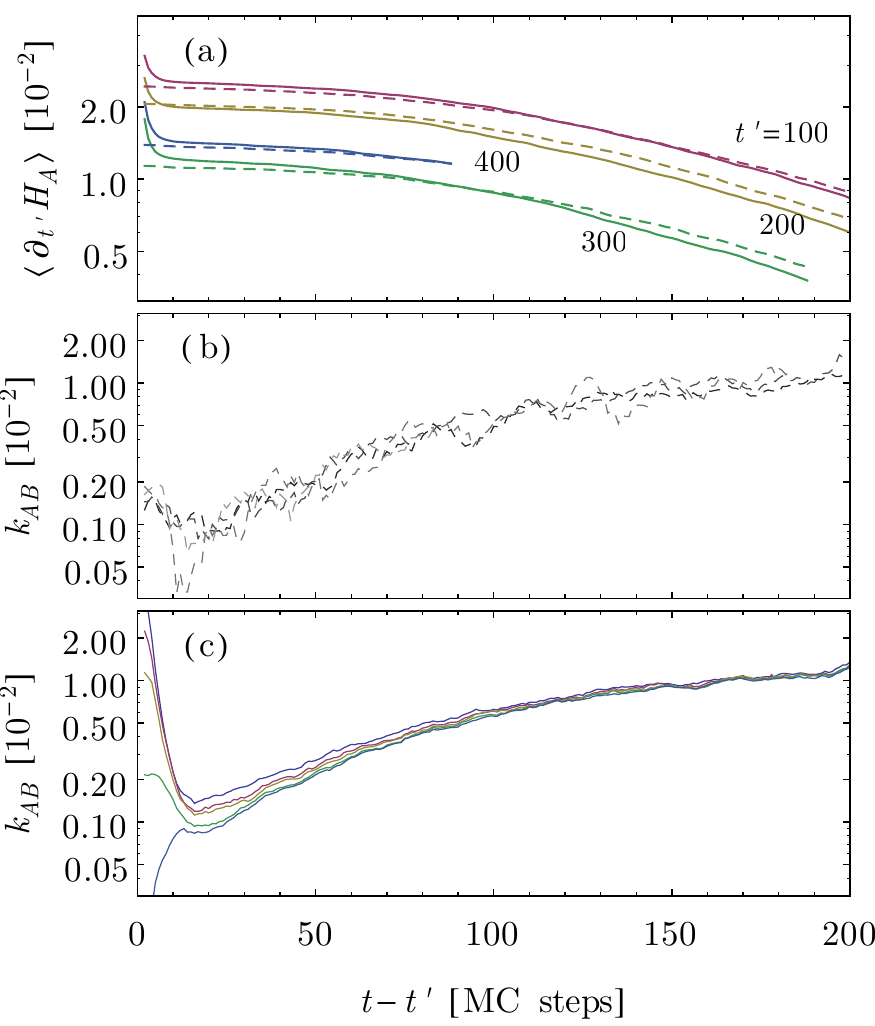}
  \caption{Time-dependent survival probabilities and rate kernels for
   the model system from Fig.~\ref{fig:forcetraj}. (a) The joint
   influx and survival probability $\Avg{\partial_{t'}H_A(t,t';\Delta
     t)}=P_A(t;t';\Delta t)=q_{BA}(t')S_A(t|t')$ as a function of $t-t'$, shown
   for different values of $t'$ ($t'$=0,100,200,400, as indicated) and
    for $\Delta t=0,10$ (solid and dashed lines, respectively).
   $P_A(t;t';\Delta t)$ is the probability that the system enters $A$ at time
   $t'$ and still is in $A$ at $t$. It decreases monotonically
   with $t-t'$, yet more slowly when the grace interval $\Delta
   t=10$ is finite; but is non-monotonic in $t'$, as explained
   in the text.  (b) The rate kernel $k_{AB}(t|t';\Delta t)$ as a
   function of $t-t'$, for different values of $t'$, and for $\Delta
   t=5$. It is seen that the rate kernels for the different times $t'$
   collapse. (c) The rate kernels $k(t-t';\Delta t)$ averaged over
   $t'$, for chosen time lags $\Delta t=0,2,4,10,20$, from top to
   bottom. It is seen that for $t-t'>\Delta t$ and $\Delta
   t>\tau\ts{trans}\approx 10$, the rate kernels become independent of
    $\Delta t$. The macroscopic rate kernel can be identified with this
    limiting function.}
  \label{fig:kPanel}
\end{center}
\end{figure}

These observations are illustrated in Fig.~\ref{fig:kPanel}, for the
clock-resetting model system introduced earlier. We recall that the quantity
$\Avg{\partial_{t'}H_A(t,t';\Delta t)}=P_A(t;t';\Delta t)$ measures the
probability that the system enters $A$ at time $t'$ and stays in $A$ at least
until $t$; it is thus given by the influx propensity $q_{BA}(t';\Delta t)$ at
$t'$ times the survival probability in $A$, $S_A(t|t';\Delta t)$:
$P_A(t;t';\Delta t)=q_{BA}(t';\Delta t)S_A(t|t';\Delta t)$. 
Fig.~\ref{fig:kPanel}a shows that $P_A(t,t';\Delta t)$ decreases with $t-t'$
because $S_A(t|t';\Delta t)$ decreases with $t-t'$. $P_A(t;t';\Delta t)$ is
non-monotonic as a function of $t'$, reflecting the fact that the flux
$q_{BA}(t';\Delta t)$ from $B$ into $A$ shows oscillations
 as the system equilibrates---these arise because the probability that the
system is in $B$ relaxes in an oscillatory fashion, see
Figs.~\ref{fig:forcetraj}, \ref{fig:HXplot}.

The rate kernel $k_{AB}(t|t';\Delta t)$ is defined as the flux of trajectories
that enter $A$ at time $t'$ and leave $A$ for the first time at time $t$
divided by the flux of trajectories that enter $A$ at time $t'$ and remain in
$A$ till time $t$, see Eqs.~\ref{eq:kdefstrict}, \ref{eq:kdefgrace}. This
conditioning removes the dependence on the start time $t'$ if the system is
time-homogeneous. Fig.~\ref{fig:kPanel}b shows that $k_{AB}(t|t';\Delta t)$
indeed depends only on the time difference: $k_{AB}(t|t'; \Delta
t)=k_{AB}(t-t';\Delta t)$, so that we can conclude that our model system is
indeed time homogeneous. By the same token, we may now average data for the 
same $t-t'$ but different $t'$, leading to the result shown in
Fig.~\ref{fig:kPanel}c. It is seen that $k_{AB}(t-t';\Delta t)$ starts out with a high crossing rate, passes
through a minimum and then rises to saturate at a constant value. The initial
drop is due to the rapid correlated recrossings of the diving surface
$q_A=q_B=q^*$; importantly, the detection of these rapid recrossings is
strongly suppressed when the grace interval $\Delta t$ is chosen to be larger
than the microscopic relaxation time $\tau\ts{trans}$, $\Delta
t\geq\tau\ts{trans}\approx 10$, as seen in panel c.  The subsequent minimum in
$k_{AB}(t-t';\Delta t)$ corresponds to local relaxation of the system inside
the basin of the new macroscopic state. The rate kernel remains small as long
as the force that tends to flip the system back to the other state on a time
scale $\tau_\phi$ is still low. Finally, as the switching force increases with
time, the rate $k_{AB}$ increases, to reach a plateau value when the switching
force becomes constant.  Fig.~\ref{fig:kPanel}c also shows that
$k_{AB}(t-t';\Delta t)$ becomes independent of $\Delta t$ when $\Delta
t>\tau\ts{trans}\approx 10$ (and $t-t'>\Delta t)$. This is crucial:
it allows us to define the macroscopic rate function $k_{AB}(t-t')$, and it
justifies \emph{a posteriori} the two-state description with a rate kernel that
depends on the time since the last switching event.

\section{Discussion}\label{sec:discussion}

It is instructive to compare the microscopic expression for the rate
kernel in a non-Markovian system, Eq.~\ref{eq:nonMarkovidentifydelta},
with that for the rate constant in a Markovian system,
Eq.~\ref{eq:k_AB_2}. Both expressions count only those switching
events where the system goes from one macroscopic state to the other
over a time $2 \Delta t$.  For a Markov system in stationary state,
the recrossing flux $j_{AA}\propto \Avg{h_A(t-\Delta
  t)\dot{h}_B(t)h_A(t+\Delta t)} = 0$, as discussed above (see also
Appendix \ref{sec:jAA}). That is, if we chose to count recrossings in
the microscopic expression for the rate function instead of using
Eq.~\ref{eq:k_AB_2}, this would not change the net result except for
additional noise. In fact, the Bennett-Chandler expression
$\Avg{\dot{h}_B(t)}_{A_0}=
\Avg{\dot{q}(0)\delta(q(0)-q^*)\theta(q(t)-q^*)}/\Avg{h_A}$
also counts trajectories that come from $B$, cross and recross the
dividing surface (a number of times), and then go to $B$; however,
their net contribution to the average is indeed zero
\cite{chandler78}.

For a non-Markovian system,  $\Avg{h_A(t-\Delta t) \dot{h}_B(t) h_A(t+\Delta
t)}$ may be non-zero even in stationary state, due to the presence of
memory over $\Delta t$. In contrast to the Markov case, here these transient
recrossings should be explicitly excluded, for the reasons already mentioned
above: If we take the macroscopic rate function to be defined as the propensity
to switch, given the time that has passed since the last macroscopic switching
event, then we should indeed only count the truly macroscopic switching events
in Eq.~\ref{eq:nonMarkovidentifydelta}.

Whether a switching event counts as a macroscopic switching event ultimately
depends on the chosen macroscopic time resolution $\Delta t$. We can define a
macroscopic switching event in an unambiguous manner if the time scale for
recrossing $\tau\ts{trans}$, is well separated from the slow memory time scale
$\tau\ts{slow}$ of the system, so that $k(t|t';\Delta t)$ is independent of
$\Delta t$ in the regime $\tau\ts{trans}<\Delta t<\tau\ts{slow}$. It may happen
that the time scales $\tau\ts{trans}$ and $\tau\ts{slow}$ are not well
separated; then, to match the experimentally measured rate constant, it becomes
critical to tune $\Delta t$ exactly to the experimental time resolution.
However, the rate kernel defined in this way will then depend on the details
of the experiment at least over some range of residence times $t-t'$, which
limits its use as a description of intrinsic properties of the system. Thus in
a non-Markovian system described by Eq.~\ref{eq:dPBdt_NM}, a time scale
separation between rapidly decaying transient memory and a well-defined slow
memory time scale is again essential.

In this context, it is noteworthy that we have used conventional indicator
functions $h_A(t)$ and $h_B(t)$, instead of the indicator functions $h_{\cal
A}(t)$ and $h_{\cal B}(t)$ as introduced by Van Erp and coworkers \cite{erp03}.
The latter are defined such that $h_{\cal
  A}(t)$ is 1 if the system was more recently in $A$ than in $B$,
and vice versa. The advantage of those indicator functions is that by a
judicious choice of the dividing surfaces $q_A$ and $q_B$, only the macroscopic
switching events are counted, i.e.~the crossing events where the system goes
from one basin of attraction to the other \cite{erp03}. If a clear separation
exists between the transient time scale and the slow time scale, and
macroscopic switching events can thus be defined uniquely (as discussed above),
then the indicator functions $h_{\cal A,B}$ are useful also in the framework
presented here. For example, these indicator functions would then obviate the
need to introduce a grace interval, as we have done for the non-Markovian
system. The advantage of the conventional indicator functions is, however, that
the corresponding correlation functions will inform us whether there exists
such a clear separation of time scales: if so, the correlation functions
exhibit a plateau, from which the macroscopic rate constant can be obtained.

Finally, now that we have derived microscopic expressions for
macroscopic rate constants for both time-inhomogeneous Markov systems
and non-Markov systems, the question arises whether these expressions
can also be evaluated efficiently in a computer simulation. In the
accompanying paper \cite{becker12a} we present a new numerical technique that
makes this possible.\\

\subsection*{Acknowledgments} We thank Daan Frenkel, Peter Bolhuis and David
Chandler for many useful discussions. This work is part of the research program
of the ``Stichting voor Fundamenteel Onderzoek der Materie (FOM)'', which is
financially supported by the ``Nederlandse organisatie voor Wetenschappelijk
Onderzoek (NWO)''.

\appendix

\section{Vanishing recrossing fluxes}\label{sec:jAA}

We consider the terms $j_{XX}(t)=\Avg{h_X(t-\Delta t)\dot{h}_B(t)h_X(t+\Delta
t)}$ appearing in Eq.~\ref{eq:Cdot_4}, where $X=A,B$. Noting that (a) for
$q_A=q_B$ we have $\dot{h}_B(t)+\dot{h}_A(t)=0$, we can restrict our attention
to the quantity $\tilde{\jmath}(t)=\Avg{h_X(t-\Delta t)\dot{h}_X(t)h_X(t+\Delta
t)}$.

Assuming memory loss over time intervals $\Delta t$ (b), and regarding
$t^\pm=t\pm\Delta t$ as fixed, we can factor the joint probabilities as
\begin{eqnarray}
\tilde\jmath(t) &=& \partial_t \Avg{h_X(t^-) h_X(t)h_X(t^+)}\nonumber\\
&=&  \partial_t (\Avg{h_X(t^-)}\Avg{h_X(t)}_{X_{t^-}}
\Avg{h_X(t^+)}_{X_t})\nonumber\\ 
&=& \Avg{h_X(t^-)} \partial_t
\Avg{h_X(t)}_{X_{t^-}} \Avg{h_X(t^+)}_{X_t}. 
\end{eqnarray}
We further assume (c) that the system is time-homogeneous over the time
$\Delta t$, in other words, any external driving, or background relaxation of
slow degrees of freedom, happens much slower. Then we may rewrite the
conditional probabilities as functions of the time difference,
\begin{equation}
\Avg{h_X(t)}_{X_{t'}} = P_{X|X}(t-t'). 
\end{equation}
Then by the product rule
\begin{eqnarray}\label{eq:condDelta}
\tilde\jmath(t) &\propto& \partial_t P_{X|X}(t-t^-) P_{X|X}(t^+-t)\nonumber\\
&=& \dot P_{X|X}(\Delta t) P_{X|X}(\Delta t) - P_{X|X}(\Delta t)
\dot P_{X|X}(\Delta t)\nonumber\\
&=&0.
\end{eqnarray}
By consequence both recrossing fluxes $j_{XX}$ in Eq.~\ref{eq:Cdot_4} vanish. 

Importantly, this conclusion rests on the three assumptions of (a) a transition
region with negligible  occupancy, (b) a time lag $\Delta t$ long enough for
any fast internal memory of the system to decay, and (c) at the same time short
enough so that the system state is quasi-stationary with respect to any
external driving or other slow relaxation dynamics.

\section{Influence of the macroscopic time scale}

\begin{figure}[htp]
\begin{center}
 \includegraphics[width=7cm]{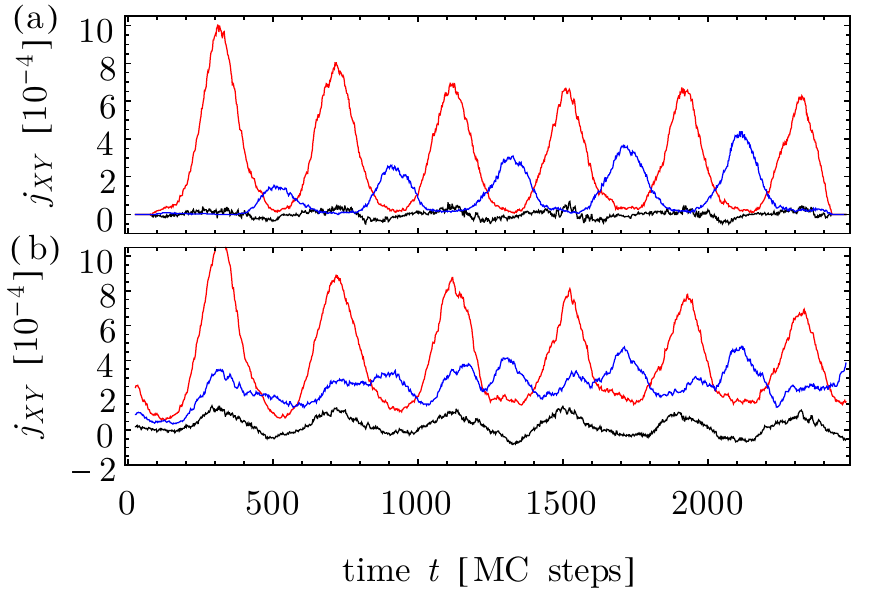}
  \caption{Variants of Fig.~\ref{fig:jABforced}b, 
  with $\Delta t$ set to 100 (a) or 2 (b). In either case, the recrossing flux
  does not vanish; in (a) the force changes noticeably during $\Delta t$, while
  in (b) the system retains memory on the scale $\Delta t$, which is seen
  clearly in the higher harmonic components in $j_{BA}$. Rate functions are
  not given by Eq.~\ref{eq:k_AB_M}.}
  \label{fig:jlongshort}
\end{center}
\end{figure}

Fig.~\ref{fig:jlongshort} is analogous to Fig.~\ref{fig:jABforced}b in the main
text, but for choices of time lag $\Delta t$ that are not well separated from
either microscopic relaxation or external driving; the re-crossing flux
$j_{AA}$ is seen to deviate from 0.

\section{Solution for the renewal rate equations}\label{sec:nonmarkovmacrosol}




The rate equations \ref{eq:dPBdt_NM} in the non-Markovian clock-resetting model
may be rewritten in Matrix form as
\begin{gather}
\vec{q}(t)  = 
\int_{0}^{t}
-\partial_{t}\mat{S}(t|t')\vec{q}(t')\d t'
-\partial_{t}\tilde{\mat{S}}(t)\vec{p}(0), \label{eq:nmconv}\\
\text{ where}\nonumber\\
\mat{S}(t|t')  =  \left[\begin{array}{cc}
0 & S_A(t|t')\\
S_B(t|t') & 0\\
\end{array}\right],\;
\tilde{\mat{S}}(t)=\left[\begin{array}{cc}
 \tilde{S}_A(t) & 0\\
 0 & \tilde{S}_B(t)
\end{array}\right],
\nonumber\\
\vec{q}(t)  =  \left[\begin{array}{c}
q_{AB}(t)\\
q_{BA}(t)
\end{array}\right],\text{ and }
\vec{p}(t) = \left[\begin{array}{c}
P_A(t)\\
P_B(t)
\end{array}\right].
\nonumber
\end{gather}
The net probability flux appearing on the left hand side of
Eq.~\ref{eq:dPBdt_NM} is then recovered as 
$\partial_t P_B(t)=-\partial_tP_A(t)=q_{AB}(t)-q_{BA}(t)$. We are concerned
with the time-homogeneous case only, so that we may write
$\mat{S}(t|t')=\mat{S}(t-t')$.

The macroscopic pre-history for $t<0$ in Eq.~\ref{eq:nmconv} is summarized in
the $\tilde{\mat S}\vec p$ term. More precisely,
\begin{equation}
-\partial_{t}\tilde{\mat{S}}(t)\vec{p}(0) =
\int_{-\infty}^0-\partial_{t}\mat{S}(t-t')\vec{q}(t')\d t'
\end{equation}
describes the first-switch propensities of trajectories that started in states
$A,B$ with probabilities $\vec{p}(0)$ at $t=0$ (having arrived at earlier
times). In the case that the system was held in a stationary state for all 
times $t'<0$, it is possible to show that 
\begin{equation}
\tilde{S}_X(t)=\frac{\int_{-\infty}^{0}S_X(t-t')\d
t'}{\int_{-\infty}^{0}S_X(-t'')\d t''}.\label{eq:specsol}
\end{equation}

To write a closed-form solution to Eq.~\ref{eq:nmconv}, note that this is a
convolution integral. After Laplace transformation ($f(s)=\int_0^\infty
e^{-st}f(t)$) we may rewrite the equation as 
\begin{eqnarray}
\vec{q}(s) &=& [-s\mat{S}(s)+\mat{S}(t=0)]\vec{q}(s) \\
&&+ [-s\tilde{\mat{S}}(s)+\tilde{\mat{S}}(t=0)]\vec{p}(t=0)\nonumber\\
&=& [-s\mat{S}(s)+\bar{\mat1}]\vec{q}(s) + 
[-s\tilde{\mat{S}}(s)+\mat1]\vec{p}(t=0),\nonumber
\end{eqnarray}
where
\begin{eqnarray}
\tilde{\mat{S}}(t=0) &=& \mat1  =  \left[\begin{array}{cc}
1 & 0\\
0 & 1
\end{array}\right] \text{ and }
\mat{S}(t=0) = \bar{\mat1}  =  \left[\begin{array}{cc}
0 & 1\\
1 & 0
\end{array}\right].\nonumber
\end{eqnarray}
We may rearrange to give an explicit solution to the
clock-resetting rate equation in Laplace space,
\begin{equation}\label{eq:expsol}
\mathbf{q}(s) =  
\left[\mat1-\bar{\mat1}+s\mat{S}(s)\right]^{-1}
[\mathbf{1}-s\tilde{\mat{S}}(s)]\mathbf{p}(0).
\end{equation}
Here the transient survival matrix $\tilde{\mat{S}}$ is either given as part of
the specification of the problem or may be obtained from $\mat S$ via
Eq.~\ref{eq:specsol} in the case of stationary pre-history. 

The stationary survival matrix
$\mat S$ is given as the solution of the equation
\begin{gather}\label{eq:Sfromk}
\partial_t\mat S(t|t') = -\mat k(t|t')\mat S(t|t'); \; \mat
S(t'|t')=\bar{\mat1},\\ \text{where }\mat k(t|t') = 
\left[\begin{array}{cc}
k_{AB}(t|t') & 0\\
0 & k_{BA}(t|t')
\end{array}\right],
\end{gather}
which is just given by 
$\exp\bigl[-\int_{t'}^{t} \mat{k}(t''|t')\d t''\bigr] \bar{\mat1}.
$
\section{Definition of the uninterrupted indicator function $H$}
\label{sec:Hmath}

When writing the function $H_A(t,t')$ as a continuous product, some
purely mathematical subtleties arise. These are immaterial for the
physics of two-state switching, and disappear when discretizing time as would
effectively be done in either experiments or simulation.

Without restriction, we take the state indicator functions $h_X(t)$ to be
right-continuous; i.e.~we impose that $h_X(t)=h_X(t_+)=\lim_{\epsilon\searrow
0}h_X(t+\epsilon)$. This means in particular that the time derivatives are
pre-point delta functions at the switching times: $\dot h_X(t)=\pm
\delta(t-t_{\mathrm{switch}\,-})$. It is then convenient to define
\begin{equation}
H_X(t,t')\equiv \prod_{t>t''\geq t'} h_X(t'');
\label{eq:Hdefrightcont}
\end{equation}
this definition makes $H_X(t,t')$ left-continuous in $t$ and right-continuous
in $t'$: $H_X(t_-,t'_+)=H_X(t,t')$. Then the relations
Eq.~\ref{eq:dot_H_X} given in the main text, 
\begin{gather}
\partial_t H_X(t,t') = \dot h_X(t)H_X(t,t')\leq 0\text{ and }\nonumber\\
\partial_{t'} H_X(t,t') = H_X(t,t')\dot h_X(t') \geq 0,\label{eq:Hdotrigorous}
\end{gather}
are justified. Other conventions for the sample paths are possible but may
require the insertion of limits in Eq.~\ref{eq:dot_H_X}.

\bibliographystyle{unsrt}

\end{document}